\documentclass[10pt]{article}
\usepackage{amsmath}
\usepackage{amssymb}
\usepackage{natbib}
\usepackage{graphicx}

\begin{document}
\bibliographystyle{plainnat}
\setcitestyle{numbers,square}

\title{\Large{A PARALLEL CODE FOR MULTIPRECISION \\
              COMPUTATIONS OF THE LANE--EMDEN \\ 
              DIFFERENTIAL EQUATION}}

\author{Vassilis S. Geroyannis$^1$, Vasileios G. Karageorgopoulos$^2$ \\
$^{1,2}$Department of Physics, University of Patras, Patras, Greece \\
\textit{$^1$vgeroyan@upatras.gr, $^2$vkarageo@upatras.gr}}

\maketitle

\begin{abstract}
We compute multiprecision solutions of the Lane--Emden equation. This differential equation arises when introducing the well-known polytropic model into the equation of hydrostatic equilibrium for a nondistorted star. Since such multiprecision computations are time-consuming, we apply to this problem parallel programming techniques and thus the execution time of the computations is drastically reduced. 
\end{abstract}

\section{Introduction}
The aim of this study is to compute multiprecision solutions of the Lane--Emden equation of stellar hydrodynamics by a code implementing the Runge--Kutta--Fehlberg method of fourth and fifth order (see e.g. \citep{EU96}, Sec. 17.3.4.4; see also \citep{GV12}, Sec.~2), working in the multiprecision environment of the ``Fortran--90 Multiprecision System'', abbreviated MPFUN90, written by D.~H.~Bailey (\citep{B95a, B95b} and references therein) --- available in http://crd-legacy.lbl.gov/ $\sim$dhbailey/mpdist/, and licensed under the Berkeley Software Distribution License found in that site. Such highly accurate solutions can be used for checking other numerical codes and prescribing a measure of their accuracy.

Since however multiprecision computations are time-consuming, we proceed with applying parallel programming techniques, appropriate for multicore machines. The Open Multi-Processing (OpenMP; http//openmp.org/) is an Application Program Interface (API) supporting shared-memory parallel programming in C/C++ and Fortran. Several compilers implement the OpenMP API. The GNU Compiler Collection (GCC; http://gcc.gnu.org/) includes a Fortran--95 compiler, so-called ``gfortran'' (http://gcc.gnu.org/fortran/), licensed under the GNU General Public License (GPL; http://www.gnu.org/licenses/gpl.html); the official manual of gfortran \citep{GF} will hereinafter be referred to by the abbreviation ``GF-M''. The GCC 4.7.x releases, including corresponding gfortran 4.7.x releases, support the OpenMP API Version 3.1; the official manual of this version \citep{OMP} will hereinafter be referred to by the abbreviation ``OMP-M''. This OpenMP version is used here with gfortran.

\section{Stellar Polytropic Models}

Polytropic models have been widely applied to the study of nondistorted (hence spherical) stars. Such stars obey the equations of hydrostatic equilibrium,
\begin{equation}
\frac{dP}{dr}=-\frac{Gm\rho}{r^2}, \qquad \qquad 
\frac{dm}{dr}=4\pi r^2\rho,
\label{eq1}
\end{equation}
where $P(r)$ is the pressure, $m(r)$ the mass inside a sphere of radius $r$, and $\rho(r)$ the density. As ``equation of state" (EOS) in the polytropic models of stars is used the well-known ``polytropic'' EOS (\citep{CH39}, Chapter~IV, Eq.~(1))
\begin{equation}
P=K\rho^{1+(1/n)} .
\label{eq2}
\end{equation}
We can introduce the so-called ``normalization equations'' (\citep{CH39}, Chapter~IV, Eqs.~(8a) and (10a), respectively)
\begin{equation}
\rho = \lambda \, \theta^n, \qquad r = \alpha \, \xi,
\label{eq3}
\end{equation}
where $\lambda$ is the central density of the star, $\lambda = \rho_\mathrm{c}$, and $\alpha$ a length defined by (\citep{CH39}, Chapter~IV, Eq.~(10b))
\begin{equation}
\alpha=\left[\frac{(1+n)K\lambda^{(1/n)-1}}{4\pi G}\right]^{1/2}.
\label{eq4}
\end{equation}
Accordingly, using $\lambda$ as the ``polytropic unit of density'' and $\alpha$ as the ``polytropic unit of length'', we verify that, in such ``polytropic units'', $\theta^n$ is the measure of density and $\xi$ the measure of length.
 
We then insert Eqs.~\eqref{eq2} and \eqref{eq3} into Eqs.~(\ref{eq1}a,\,b) and, after some algebra, we obtain the so-called ``Lane--Emden equation" (cf. \citep{CH39}, Chapter~IV, Eqs.~(11) and (12))
\begin{equation}
\theta'' + \, \frac{2}{\xi} \, \theta' = \, -\theta^n, 
\label{eq5a}
\end{equation}
which, when integrated along a prescribed integration interval,
\begin{equation}
\xi \in [\xi_\mathrm{start} = 0, \, \xi_\mathrm{end}] = 
\mathbb{I}_\xi \subset \mathbb{R},
\label{eq5I}
\end{equation}
with initial conditions
\begin{equation}
\theta(\xi_\mathrm{start}) = 1, \qquad \qquad \theta'(\xi_\mathrm{start})=0
\label{eq5bc}
\end{equation}
where primes denote derivatives with respect to $\xi$, gives as solution the so-called ``Lane--Emden function" $\theta[\mathbb{I}_{\xi}\subset\mathbb{R}$]. 

The second-order differential equation~\eqref{eq5a} together with the initial conditions~\eqref{eq5bc} establish an ``initial value problem'' (IVP). Our aim in this study is to compute multiprecision solutions $\theta[\mathbb{I}_\xi]$ of this IVP.
There are, however, two problems regarding Eq.~\eqref{eq5a}. First, to remove the indeterminate form $\theta'/\xi$ at the origin, appearing in the left-hand side, we modify the denominator by adding a tiny quantity, $\xi_0$, to it, i.e. $\theta'/(\xi+\xi_0)$. Since $\xi_0$ is small, the initial conditions~\eqref{eq5bc} are valid at the starting point $\xi_\mathrm{start} + \xi_0 = \xi_0$ as well. Accordingly, the integration interval becomes
\begin{equation}
\xi \in [\xi_0, \, \xi_\mathrm{end}] = 
\mathbb{I}_{\xi0} \subset \mathbb{R}.
\label{eq5Imod}
\end{equation}

The second problem is that for values $\xi$ greater than the first root $\xi_1$ of $\theta(\xi)$, $\xi>\xi_1$, the Lane--Emden function changes sign, $\theta(\xi)<0$, and thus the term $\theta^n$ in Eq.~\eqref{eq5a} becomes undefined --- raising a negative real number to a real power, e.g. $(-2.7)^{1.5}$, is not defined in $\mathbb{R}$. This undefined issue can be removed by taking instead the real power of the absolute value of $\theta$, $|\theta|^n$. Note however that this ``numerical trick" is appropriate only when interested in finding the first root $\xi_1$ of the function $\theta$; while it becomes unreliable when searching for higher roots of $\theta$ (i.e. $\xi_2$, $\xi_3$, $\xi_4$, \dots). To compute such higher roots --- involved, e.g., in models of planetary systems --- we need to apply the so called  ``complex--plane strategy"; full details of this method can be found in \citep{GK14}.

As it is usual in numerical analysis, we proceed by transforming Eq.~\eqref{eq5a} into a system of two first-order differential equations, with the IVP under consideration having then the form
\begin{equation}
\theta'=\eta, \quad \eta'= - \, \frac{2}{(\xi+\xi_0)} \, \eta - |\theta|^n, \quad \xi\in\mathbb{I}_{\xi}, \quad \theta(0)=1, \quad \eta(0)=0.  
\label{eq7}
\end{equation}
By solving this IVP, we can compute several significant physical characteristics of a stellar model with finite radius (i.e. $n \in [0,\ 5)$, since the model with polytropic index $n=5$ has an infinite radius). If the radius $R$, the mass $M$, and an appropriate polytropic index $n$ are known for a stellar model, then the polytropic units $\alpha$ and $\lambda$ are given by 
\begin{equation}
\alpha = \frac{R}{\xi_1(n)}, \qquad \qquad 
\lambda=-\left[\frac{\xi_1(n)}
        {4\pi\theta'(\xi_1(n))}\right]\frac{M}{R^3},  
\label{eq8}
\end{equation}
where the symbol $\xi_1(n)$ shows that the root $\xi_1$ varies with $n$.

The ratio of the central density, $\lambda = \rho_\mathrm{c}$, to the mean density, $\bar{\rho}$, is given by (cf. \citep{CH39}, Chapter~IV, Eq.~(78))
\begin{equation}
\frac{\lambda}{\bar{\rho}} = 
- \, \frac{\xi_1(n)}{3\, \theta'(\xi_1(n))} \ .  
\label{eq12}
\end{equation}

Furthermore, it can be proved that the coefficient $N_n$ appearing in the mass-radius relation (\citep{CH39}, Chapter~IV, Eqs.~(72) and (75)) is equal to
\begin{equation}
N_n = \, \frac{1}{n + 1} \, \left[ \, \frac{4 \, \pi}
{_0\omega_n^{n-1}} \, \right]^{1/n},  
\label{eq13}
\end{equation}
where the coefficient $_0\omega_n$ is defined by (\citep{CH39}, Chapter~IV, Eq.~(73)) 
\begin{equation}
_0\omega_{n}= \,  -\xi_1(n)^{(n+1)/(n-1)} \, \theta'(\xi_1(n)).  
\label{eq14}
\end{equation}

Finally, the coefficient $W_n$ appearing in the central-pressure relation 
(\citep{CH39}, Chapter~IV, Eq.~(80)) is defined by (\citep{CH39}, Chapter~IV, Eq.~(81))
\begin{equation}
W_{n}= \, \frac{1}{4 \, \pi \, (n+1) \left[ \theta'(\xi_1(n)) \right]^2}.  
\label{eq15}
\end{equation}

\section{Parallel Programming}
\label{PCCE}
Our computer comprises an Intel\textsuperscript{\textregistered} Core\texttrademark \, i7--950 processor with four physical cores. This processor  possesses the Intel\textsuperscript{\textregistered} Hyper-Threading Technology, which delivers two processing threads per physical core. gfortran has been installed in this computer by the TDM-GCC ``Compiler Suite for Windows'' (http://tdm-gcc.tdragon.net/; free software distributed under the terms of the GPL).

According to GF-M (Sec.~6.1.16), to enable the processing of the OpenMP directive sentinel \texttt{!\$OMP} (OMP-M, Sec.~2.1), gfortran is invoked by the ``-fopenmp'' option. Then all lines beginning with the sentinels \texttt{!\$OMP} and \texttt{!\$} (OMP-M, Sec.~2.2) are processed by gfortran.  

We develop here a parallel programming code for solving the polytropic IVP~(\ref{eq7}). The code consists of two parts. The task of the first part is to provide all the available computer cores with the required variables. The second part performs numerical computations in parallel for all the polytropic indices given. We use the work-sharing constructs of OpenMP package (OMP-M, Sec.~2.5) in order to share the numerical work and to activate the available computer cores. 

A critical step in parallel programming is the demarcation of the shared memory by using data-sharing attribute clauses (OMP-M, Sec.~2.9.3) like \texttt{SHARED}, \texttt{PRIVATE} and \texttt{FIRSTPRIVATE} (for an indicative list of variables categorized this way, see Table~\ref{ta1}). Variables with values to be shared among all threads are declared as \texttt{SHARED}. On the other hand, variables are declared as \texttt{PRIVATE} when each thread has to work with its own copy of values of such variables. Furthermore, variables initialized according to values assigned prior to entry into parallel processing are declared as \texttt{FIRSTPRIVATE}. The computations are distributed over all the threads by the \texttt{SCHEDULE(DYNAMIC)} clause (OMP-M, Sec.~2.5.1, Table~2-1). Thus, once a particular thread finishes its allocated iteration, it returns to get a next one from the iterations that are left to be processed. The worksharing construct \texttt{DO} (OMP-M, Sec.~2.5.1) is used in order for the computer to share among its threads the work of calculating the first root for each polytropic index. The user has to provide a value to an integer variable \texttt{NMODEL}, equal to the number of threads available in his/her computer; thus the number of the models, which are to be processed in parallel, must be less or equal to \texttt{NMODEL}. 


\section{Multiprecision Environment and \\ Modified 
Runge--Kutta--Fehlberg 5(4) Code}
\label{RKF}
The computations are performed in the high precision environment of the package MPFUN90 \citep{B95a,B95b}. This package supports a flexible, arbitrarily high level of numeric precision --- with hundreds or thousands of decimal digits. MPFUN90 is written in Fortran--90. An intermediate translating program is not required, since translation of the code in multiprecision is accomplished by merely utilizing advanced features of Fortran--90. In this study, our computations are carried out with a presicion of 64 decimal digits. 

The code \texttt{DCRKF54}, developed and used in \citep{GV12}, can solve complex IVPs. We have modified properly this code in order for the modified one, so-called \texttt{DDRKF54}, to solve real IVPs in the multiprecision environment of MPFUN90. The header of the subroutine \texttt{DDRKF54} has the form
\begin{verbatim}
! Part #[000]: Header of DDRKF54
      SUBROUTINE DDRKF54(IMDL,A,B,N,Y,DEQS,H,HMIN,HMAX,
     &                                     ATOL,RTOL,QLBD,NFLAG)
\end{verbatim}
We assign to these arguments the values $\mathtt{IMDL} = n$ (integer), $\mathtt{A} = r_\mathrm{start}$ (real of proper kind), $\mathtt{B} = r_\mathrm{end}$ (real of proper kind), $\mathtt{N} = 2$ (integer), $\mathtt{Y(1)} = \theta(\xi_\mathrm{start})$ (real of proper kind), $\mathtt{Y(2)} = \theta'(\xi_\mathrm{start})$ (real of proper kind). The remaining arguments are discussed in \citep{GV12} (Appendix).

We compute the first root of the Lane--Emden function $\theta$ by combining \texttt{DDRKF54} with a code that mimics the well-known bisection algorithm. 
The basic idea is that, when the density changes sign from plus to minus, then the code returns to the last computed point, reduces the stepsize to its half, and repeats the computation at the new point. If the density is again negative, then the process is repeated.

\section{Numerical Results}
Intergration takes place along two successive intervals. The first interval is a very short one, $\mathbb{I}_1 = [10^{-26}, \ 30\times10^{-26}]$, in order for the code \texttt{DDRKF54} to be accurately initiated. The stepsize along this interval is kept constant and very small, of order  $10^{-29}$ (Table \ref{ta2}).  

The second integration interval, $\mathbb{I}_2$, extends from the end of $\mathbb{I}_1$ up to a value $\sim \frac{3}{2} \, \xi_1$. In this study, we resolve 9 polytropic models with indices $n=$ 0.00, 1.00, 1.50, 2.00, 2.45, 2.50, 3.00, 3.23, 3.25, and 3.50.  

The multiprecision results of this study are given in Tables \ref{ta3}--\ref{ta5}. A polytropic index appropriate for verifying the accuracy of our code is $n=1$, since, as it is well-known, this case has an analytical solution and the corresponding first root of the Lane--Emden function is $\xi_1 = \pi$ (\citep{CH39}, Chapter~IV).

\section*{Acknowledgments} 
The authors acknowledge the use of the Fortran--90 Multiprecision System \citep{B95a,B95b}.

\begin{table}
\caption{Inidicative list of variables involved in the computations.\label{ta1}}
\begin{center}
\begin{tabular}{llr} 
\hline \hline 
Name & Description & Category\\ \hline
\\ 
\texttt{POL\_INDEX} & polytropic index & \texttt{FIRSTPRIVATE}\\
\texttt{T\_END} & final end point& \texttt{FIRSTPRIVATE} \\
\texttt{T\_IN} & local start point & \texttt{FIRSTPRIVATE}\\
\texttt{T\_OUT} & local end point& \texttt{FIRSTPRIVATE} \\
\texttt{X} & independent variable of integration & \texttt{FIRSTPRIVATE} \\
\texttt{Y} & dependent variable(s) of integration & \texttt{FIRSTPRIVATE} \\
\texttt{H} & initial stepsize &  \texttt{FIRSTPRIVATE} \\
\texttt{TOLROOT} & error tolerance for computing root(s) & \texttt{FIRSTPRIVATE} \\
\texttt{YPRIME} & first derivative(s) of \texttt{Y} & \texttt{PRIVATE} \\
\texttt{FIRST\_ROOT} & first root $\xi_1$ & \texttt{PRIVATE} \\
\texttt{OMEGA\_0N} & $_0\omega_n$, Eq.~(\ref{eq14}) & \texttt{PRIVATE} \\
\texttt{COEFF\_NN} & $N_n$, Eq.~(\ref{eq13}) & \texttt{PRIVATE} \\
\texttt{COEFF\_WN} & $W_n$, Eq.~(\ref{eq15}) & \texttt{PRIVATE} \\
\texttt{NEQ} & number of first-order DEQs & \texttt{SHARED} \\  
\texttt{HMIN} & minimum allowed stepsize & \texttt{SHARED} \\
\texttt{HMAX} & maximun allowed stepsize & \texttt{SHARED} \\
\texttt{ATOL} & absolute error tolerance & \texttt{SHARED} \\
\texttt{RTOL} & relative error tolerance & \texttt{SHARED} \\
\texttt{QLBD} & error--tolerance coefficient of \texttt{DDRKF54}  & \texttt{SHARED} \\
\hline 
\end{tabular}
\end{center}
\end{table}

\begin{table}
\caption{Initial values of variables and values of parameters.\label{ta2}}
\begin{center}
\begin{tabular}{lrr} 
\hline \hline 
Name & Value in the first interval & Value in the second interval \\ \hline
\\ 
\texttt{T\_IN} & 0 & $30 \times 10^{-26}$ \\
\texttt{T\_END} & $30 \times 10^{-26}$ & model dependent \\  
\texttt{EPS} & $10^{-42}$ & $10^{-42}$ \\
\texttt{H} & $10^{-29}$ & $10^{-4}$ \\ 
\texttt{HMIN} & $10^{-10}$ & $10^{-5}$ \\
\texttt{HMAX} &  $10^{-1}$ & $10^{-1}$ \\
\texttt{RTOL} & $10^{-42}$ & $10^{-32}$ \\
\texttt{ATOL} & $10^{-42}$ & $10^{-46}$ \\
\texttt{QLBD} & $7.5 \times 10^{-1}$ & $7.5 \times 10^{-1}$\\
\texttt{TOLROOT} & inactive & $10^{-46}$ \\
\hline 
\end{tabular}
\end{center}
\end{table}

\begin{table}
\caption{Characteristics of the Lane--Emden function, I.\label{ta3}}
\begin{center}
{\footnotesize
\begin{tabular}{lcc}
\hline \hline
\\ 
$n$ & $\xi_1$ & $-\xi_1^2 \, \theta'(\xi_1)$ \\ 
\hline 
0.00 & 2.44948 97427 83178 09819 72840 74705 9 & 4.89897 94855 66356 19639 45681 49411 8 \\
1.00 & 3.14159 26535 89793 23846 26433 83279 5 & 3.14159 26535 89793 23846 26433 83279 5 \\  
1.50 & 3.65375 37362 19122 42460 90942 80459 2 & 2.71405 51201 08645 71902 45332 77696 4 \\
2.00 & 4.35287 45959 46124 67697 35700 61526 1 & 2.41104 60120 96893 78364 84427 44671 4 \\
2.45 & 5.23614 14048 69233 45598 24188 11476 5 & 2.20681 79791 33278 31713 43386 65923 3 \\ 
2.50 & 5.35527 54590 10779 45990 93600 02973 6 & 2.18719 95655 17078 95321 95209 89736 0 \\
3.00 & 6.89684 86193 76960 37545 45281 87123 1 & 2.01823 59509 66228 40281 28131 70057 9 \\
3.23 & 7.91690 48976 05477 15893 80785 74290 1 & 1.95492 90412 23479 62411 38693 00245 9 \\
3.50 & 9.53580 53442 44850 44410 47426 25789 3 & 1.89055 70934 43116 39390 85853 05853 5 \\
\hline 
\end{tabular}} 
\end{center}
\end{table}

\begin{table}
\caption{Characteristics of the Lane--Emden function, II.\label{ta4}}
\begin{center}
{\footnotesize
\begin{tabular}{lcc}
\hline \hline 
$n$ & $\lambda_s/\rho_s$ & $_0\omega_n$ \\ 
\hline 
0.00 &  1.00000 00000 00000 00000 00000 00000 0$(+0)$ & 3.33333 33333 33333 33333 33333 33333 3$(-1)$ \\
1.00 &  3.28986 81336 96452 87294 48303 33292 0$(+0)$ & 1.00000 00000 00000 00000 00000 00000 0$(+0)$ \\  
1.50 &  5.99070 45163 03533 56943 21224 39795 1$(+0)$ & 1.32384 29925 64083 05364 42661 94225 6$(+2)$ \\
2.00 &  1.14025 42861 95126 16631 26684 57624 2$(+1)$ & 1.04949 80935 71378 17586 20320 76059 5$(+1)$\\
2.45 &  2.16843 46128 86921 36359 86989 82587 2$(+1)$ & 4.13518 51814 61111 26824 04912 87756 7$(+0)$ \\ 
2.50 &  2.34064 62267 06270 85025 51279 20224 5$(+1)$ & 3.82662 30435 74210 20115 81302 79329 3$(+0)$ \\
3.00 &  5.41824 81107 34076 30151 80169 72241 2$(+1)$ & 2.01823 59509 66228 40281 28131 70057 9$(+0)$ \\
3.23 &  8.46085 06877 90592 67790 61083 54949 5$(+1)$ & 1.57926 45176 10004 22561 84996 73130 6$(+0)$ \\
3.50 &  1.52883 66297 09408 05111 21531 08344 2$(+2)$ & 1.20425 46220 02747 85274 19573 06534 3$(+0)$ \\
\hline 
\end{tabular}}
\end{center}
\end{table}

\begin{table}
\caption{Characteristics of the Lane--Emden function, III.\label{ta5}}
\begin{center}
{\footnotesize
\begin{tabular}{lcc} 
\hline \hline 
$n$ & $N_{n}$ & $W_{n}$ \\ 
\hline 
0.00 &  0.00000 00000 00000 00000 00000 00000 0$(+0)$ & 1.19366 20731 89215 01826 66282 25293 9$(-1)$ \\
1.00 &  6.28318 53071 79586 47692 52867 66559 0$(+0)$ & 3.92699 08169 87241 54807 83042 29099 3$(-1)$ \\  
1.50 &  4.24216 67959 99143 05430 08004 94342 2$(-1)$ & 7.70140 37135 09219 86069 32395 45144 2$(-1)$ \\
2.00 &  3.64747 94209 21269 20134 43373 45944 9$(-1)$ & 1.63818 33457 74008 02382 13552 61677 0$(+0)$ \\
2.45 &  3.51539 09189 35827 95514 94851 07072 2$(-1)$ & 3.56027 44648 54173 84876 18285 43047 9$(+0)$ \\ 
2.50 &  3.51500 94082 14026 96517 75801 90482 5$(-1)$ & 3.90906 17224 91728 32009 55170 68744 8$(+0)$ \\
3.00 &  3.63939 06192 87273 76108 24669 23988 2$(-1)$ & 1.10506 81023 46314 79930 43159 71504 1$(+1)$ \\
3.23 &  3.77536 15181 81869 22747 08226 81199 8$(-1)$ & 1.93379 07742 64900 15775 48009 21363 7$(+1)$ \\
3.50 &  4.01043 51975 50305 19191 13723 34788 4$(-1)$ & 4.09098 16957 57662 25263 91197 48667 2$(+1)$ \\
\hline 
\end{tabular}}
\end{center}
\end{table}

\end{document}